# Charges électriques et tribologie des isolants.
# Electrical charges and tribology of insulating materials.

Christelle GUERRET-PIECOURT[a], Sandrine BEC[b], Daniel TREHEUX[a]

[a]IFoS, UMR CNRS 5621, Ecole Centrale de Lyon, BP 163, 69131 Ecully Cedex, France
[b]LTDS, UMR CNRS 5513, Ecole Centrale de Lyon, BP 163, 69131 Ecully Cedex, France
E-mail Christelle.Guerret@ec-lyon.fr

**Résumé.** La description de la génération de charges électriques lors du contact entre isolants a été faite de longue date. En revanche, la prise en compte quantitative des effets de ces charges est rarement évoquée dans les bilans énergétiques concernant les problèmes de frottement. En nous appuyant sur des résultats publiés et sur nos propres expérimentations sur le piégeage des charges dans les isolants, nous montrons l'importance de l'action de ces charges électriques sur l'énergie d'interaction pendant le frottement. Enfin, nous proposons d'évaluer quantitativement cette contribution grâce à l'utilisation complémentaire de la méthode « miroir » et de mesures de forces de surfaces.
caractérisation diélectrique / frottement / triboélectrification / interactions interfaciales / énergie libre de surface / charge d'espace / forces de surface

**Abstract.** Electrical charges generation occurring during contact and friction of insulating materials has been identified for a long time. However the contribution of these electrical charges to the friction behaviour is usually neglected in the energetical balances. Based on published results and on our own experimental results on the ability of the dielectric materials to trap charges, we show in this study that the interaction energy during friction depends markedly on these trapped charges. Eventually, we propose the complementary use of the "mirror " method and of surface forces measurements to obtain a quantitative evaluation of this contribution.
dielectric behaviour / friction / triboelectrification / interfacial interactions / surface free energy / space charge / surface forces

## 1. Introduction

Surface forces act between two surfaces when they are in close proximity. They are very well known in the field of colloid science [1, 2] and they have also a strong impact in other fields like adhesion, wetting or friction. Resulting from very complex phenomena, these forces include mechanical, physico-chemical and electrical effects.

This paper is focused on one phenomenon which is often neglected : the effects of electrical charges trapped in the near surface or in the bulk of dielectric materials (ceramics, polymers...). In section 2, the various types of surface forces according to their range are recalled. Then, it is shown that some published results in the friction and adhesion domains suggest the major role of these electrical charges. This is followed by an overview on charge generation and trapping during contact or friction in section 3. Our own results are used in section 4 to discuss how electrical charges can influence the friction behaviour of dielectrics. In order to go further, the complementary use of the "mirror " method and of surface forces measurements is proposed in the aim to obtain a quantitative evaluation of this contribution.

## 2. Surface forces and electrical charges.

The surface forces [3] involved between two solids in contact result from the contribution of the interatomic forces between all the atoms of the tribological system. They include those of the two bodies plus those of the medium surrounding them. These surface forces can be classified according to their ranges.

### 2.1 Very-short-range forces :

Very-short-range forces result from the exchange of electrons like in covalent bonding, hydrogen bonding ... (metallic bonding is not considered in this work devoted to insulating materials).

## 2.2 Short-range forces :

Due to the various interactions between atomic or molecular electric dipoles, the van der Waals interactions are short-range forces which are the sum of three contributions : the Keesom, the Debye and the London interactions.

Different approaches have been developed to calculate the strength of the van der Waals forces. Hamaker [4, 5] has proposed a simple pairwise additive model. In the Derjaguin approximation [6], it leads for instance to the following expression for the interaction energy per unit area ($J/m^2$) for two planar surfaces :

$$W(D)=-A/12\pi D^2 \qquad (1)$$

where D is the surface separation, and A the Hamaker constant, depending on the polarizabilities of the atoms. In the case of two similar materials, with $A=10^{-19}$ J as a typical value for the Hamaker constant, the interaction energy at contact (D=0.2 nm) will be W(D=0.2 nm)=-0.066 $Jm^{-2}$. In order to include the many-body effects, Lifshitz has proposed an alternative formulation which is capable of making more precise predictions of the van der Waals force, if complete dielectric properties of the materials are available [2].

## 2.3 Long-range forces :

The coulombic interaction is the stronger and the longer ranged of the surface forces. However, this interaction is usually neglected in the energetical balances excepting for the electrical double layer interaction.

*(i) in polar environment :*

The electrical double layer interaction occurs when surfaces are immersed in a polar environment and leads to the well-known DLVO theory [7]. Direct measurements of this surface force in good agreement with the prediction of the DLVO theory have been obtained on surface force apparatus (SFA), for example between sapphire crystals in aqueous solutions [8].

*(ii) in non-polar environment :*

However, in non-polar environments (vacuum, dry nitrogen...), long range coulombic forces during friction or adhesion between two insulating materials become significant.

Horn et Smith [9, 10] have evidenced and quantified the attraction resulting from contact electrification due to non-sliding contact between smooth insulators (thin sheets of mica and silica, respectively 3.23 µm and 6.37 µm thick) in dry nitrogen atmosphere in a surface force apparatus. Surface forces and surface charges are simultaneously measured as a function of surface separation. During the first approach, the only force was the van der Waals attraction but after ten touches, a strong attractive force (several tens of mN) was measured, persisting over several micrometers, with some breaks in the data at certain separations (cf. fig. 1). Using simplifying assumptions, they gave an approximate analysis of the distance dependence of this long-range electrostatic force, from which they calculated the surface charge density per unit area. The value was ranging from 6.5 $mC/m^2$ to 10.9 $mC/m^2$, i.e. in the range of the typical value obtained during insulator-insulator contact (cf. § 3.2). The breaks were attributed to partial discharges across the gas. By integrating the electrostatic force as a function of the separation, a measurement of the work of adhesion was obtained. It varies between -6.6 $J/m^2$ and -8.8 $J/m^2$ depending on the assumption made o  the discharge phenomenon. These values are comparable to the work of cohesion of ionic and ionic-covalent solids, and much higher than values obtained for van der Waals forces . This made the authors conclude that contact electrification can lead to strong adhesion.

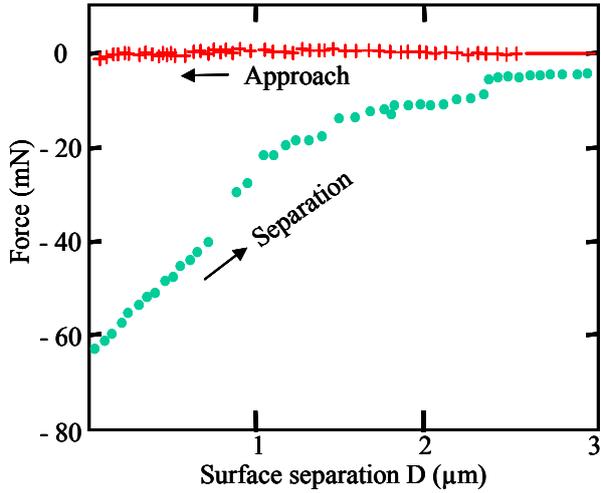

*Figure 1 :*
*From Horn R.G. and Smith D.T., Science, **256**, 362-364 (1992)*
*Force measured between crossed cylinder of mica and silica in dry nitrogen gas, as a function of separation D.*
*- Red crosses : first approach, little force measured*
*- Green circles : separation from contact (after ten touches), strong attractive force ascribed to electrostatic attraction after charge transfer between the materials. Breaks occur in the data at certain separations.*

The direct interaction in non-polar environment can be also based on the image charge effects due to the electrical charges present in the bulk of the material. Briefly, Landau and Lifshitz [11] have considered the case of a planar boundary dividing space in two regions of different dielectric constants $\varepsilon_1$ and $\varepsilon_2$ (cf. fig. 2). The effect on a charge Q in region 1 is estimated by assuming region 1 extended to fill in all space and by including interaction with a virtual image charge Q'. The magnitude of this virtual charge is related to the two dielectric constants and its position is the mirror image of charge Q considering the interface plane as a reflecting plane. Thus for a charge Q, at a distance D from the interface, the interaction energy is given by :

$$W(D) = \frac{-Q^2}{4(4\pi\varepsilon_0\varepsilon_1)D}\left(\frac{\varepsilon_2-\varepsilon_1}{\varepsilon_2+\varepsilon_1}\right) \qquad (2)$$

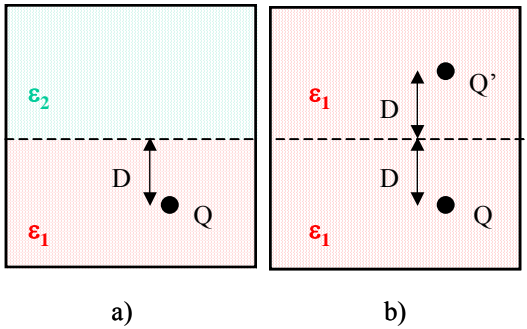

*Figure 2:*
*Illustration of the image charge interaction :*
*(a) Landau and Lifshitz [11] have considered two different dielectrics 1 and 2 (respective dielectric constants $\varepsilon_1$ and $\varepsilon_2$) and a charge Q at a distance D from the interface.*
*(b) Assuming region 1 extended to fill in all space, the charge Q experiences a force (due to the presence of dielectric 2) equal to those produced by a "virtual charge" Q' of strength $-Q(\varepsilon_1-\varepsilon_2)/(\varepsilon_1+\varepsilon_2)$ placed symmetrically to Q about the interface (i.e. at a distance 2D of Q).*

For example, consider a charge Q=1e in a medium of dielectric constant $\varepsilon_1$=10 at a distance D=0.2 nm from the boundary with a metal ($\varepsilon_2=\infty$) the image interaction energy is W(D=0.2 nm)=0.18 eV=$2.9 \cdot 10^{-20}$ J. When simply superposing $10^{19}$ charges per square meter, it gives an interaction energy of W(D=0.2 nm)=290 mJm$^{-2}$ [12].

Using this approach, Stoneham [12] has elucidated why the wetting angle for the non-reactive metal Cu on oxides is not in agreement with van der Waals theory's prediction. In this case, the electrostatic interactions were dominant. The wetting ability appeared to be related to the degree of disorder or of non-stoichiometry in the insulating material. The disorder provided high concentrations of defect charges and hence strong image interaction, which modified the interfacial energy with the metal.

Another example of the usefulness of the image charge method concerns the variation of the wetting angle of water on oxide grown on Si [12, 13]. In this case, the van der Waals forces cannot explain why the contact angle for water on the oxide surface changed from non-wetting for no-oxide to wetting for oxide of 40 Å thick. Stoneham has predicted it by assuming charged defects in the oxide and he has shown that the image terms influence interface energies, whenever there is a change in dielectric constant across the boundary. More recently, he has extended his work to polymers (polyimides, polyacetylene...) [14, 15] and has shown that the theory of defects in polymers can explain many properties of these materials, like adhesion [16, 17].

Finally, the most important concept is the idea that in some cases (notably ionic-metal systems), adhesion can be controlled by controlling charged defect populations and the distribution of carriers over near-surface traps. This latter idea has been recently validated by Stoneham and co-workers [18], who have presented modelling which showed that the radiation-induced adhesion between Ag and MgO can be well explained by the major role of interface charged defects.

In rare recent papers, some authors have taken into account the electrical charges in their experiments on friction between the two insulating materials.

Wistuba [19] has measured the change in friction and in electric charge in alumina/polytetrafluoroethylene (PTFE) sliding contact under boundary lubrication conditions. A cross-relation between the friction force and the electric charge has been observed. Both values depend on the lubrication conditions (nature, film thickness). The knowledge of the value of the electrical charge can be useful in optimising the friction conditions. Indeed, it can be a measure of the lubricant film stability or it allows the selection of the best operating conditions. This study underlines the correlation between friction force and electrical charges generated during sliding contact even if no evaluation of the direct coulombic forces has been done.

Sounilhac et al [20] have used atomic force microscopy to measure long range forces and adhesion energy between tungsten and oxide surfaces under ultrahigh vacuum. For non-stoichiometric $TiO_2$, they observed no deviation of the gradient force compared to the van der Waals theory because the sample is well conducting. For the bad conducting $W/TiO_2$ stoichiometric system, at very small tip-sample distance (below 15 nm), the force gradient versus tip-sample distance curve well fits the usual Derjaguin approximation of the van der Waals forces with $A=5.3 \cdot 10^{-19}$ J for the Hamaker constant value. For larger distance (above 15 nm), a shoulder has been observed on the force gradient curve (cf. fig.3). It has been attributed to the presence of electrostatic forces.

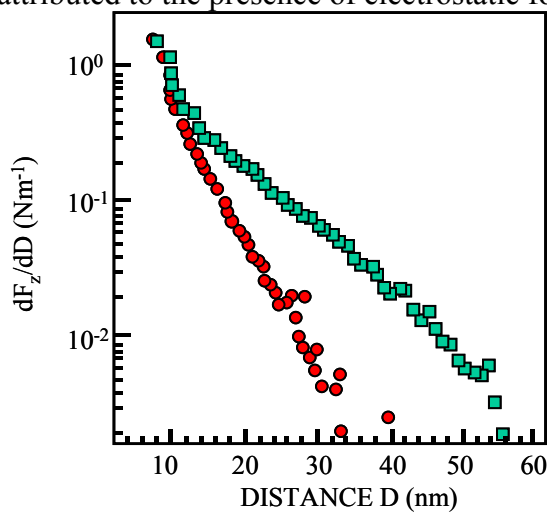

*Figure 3 :*
*From Sounilhac et al, J. Appl. Phys., 85 (1), 222-227 (1999)*

*Log-linear plot of the force gradient versus AFM tip-sample distance for the system $W/TiO_2$ [stoichiometric (green squares), and non-stoichiometric (red circles)].*
*For non-stoichiometric $TiO_2$, the data fit well respectively a retarded van der Waals law at long distance (D>15 nm) and a nonretarded van der Waals law at short distance(D<15 nm).*
*For stoichiometric $TiO_2$ at short distance (D<15 nm), the data fit a nonretarded van de Waals law. Above 15 nm the electrostatic forces dominate.*

Their estimated value of the Hamaker constant is then overestimated (the Lifshitz theory gives $A=2.9 \cdot 10^{-19}$ J), because it includes the electrostatic contribution. However, the authors could not properly described this electrostatic contribution with the existing models from the literature [21]. This can be compared with Horn et al results on surface forces between sapphire crystals in aqueous solutions [8]. They have correctly fitted their curve using the DLVO theory [22] combined, for shorter distances, with the van der Waals attraction, however the Hamaker constant used to generate the curves is about 20% higher than the calculated and expected one. Like for the $W/TiO_2$ interaction, Horn's results may be explained as following : the coulombic forces were not dominant but they could have influenced the adhesion forces and provoked this apparent increase in the Hamaker constant.

## 3. Generation and conduction of electrical charges
### 3.1 Triboemission
Charge phenomena can be linked to triboemission, (exoemission of electrons, charges particles or photons during friction [23, 24], and fractoemission [25]...). Kajdas & al [26] has proposed a complete review of these phenomena related to the boundary friction process. However, if they can be an additional source of electrical charges in the interface between the two materials, their contribution to the trapped charges in the material is not yet totally understood.

### 3.2 Triboelectrification
Triboelectrification [27] occurs when two materials are touching each other or rubbing together : electrical charges are usually transferred from one solid to the other. As recalled by Roses-Ines in his review [27], triboelectrification has been traditionally classified into three categories : metal-metal, metal-insulator, and insulator-insulator contact. The quantities and the nature of charges transferred between both materials strongly depend on the nature of the materials.

*(i)For metal-metal contact,* Harper [28] has demonstrated that electrons flow until the two Fermi levels come into coincidence. Hence, the charge transfer between two metals is proportional to their initial contact potential difference. For instance, charges for Co-Cr contact were about threefold smaller than for Au-Cr contact.

*(ii)For metal-insulator contact*, the charges acquired by an insulator from a metal depend not only on the nature of the two materials, but also on the experiment parameters (temperature, speed of sliding, environment...). Only the order of magnitude of the charge density obtained by contact electrification can be given. For organic polymers, the charge density is usually in the range of $10^{-5}$ to $10^{-3} C/m^2$ [27]. Similar charge density have been measured on $SiO_2$ and $Al_2O_3$, [29, 30].

The contact electrification may be due to the transfer of ions [31, 32] or due to material transfer. For instance, according to Kornfeld [31], insulators contain an internal electric field due to charged defects present in the lattice. So, the ions present in the atmosphere compensate this field and form a part of the surface layer. Then friction of two materials mix their surface layers. Finally, the compensation of the intrinsic field of the two bodies is disturbed and the insulator becomes electrified.

However, ion transfers is usually not considered as the dominant cause of triboelectrification [28, 33]. Actually, charges acquired by an insulator, when touched by a metal, are usually attributed to electron transfer [27]. Indeed, some authors have shown that the charge varies with metal work function [28, 34]. In this case, a logical extension of Harper's theory to conductor-insulator has been proposed by Davies. The insulator structure is considered to behave as a wide band-gap p-type semi-conductor. In a classical band structure model, the trapping levels produced by the imperfections of the material (point and structure defects) are located in this band gap [27]. However, some inconsistencies in this model (such as tunnelling distance much larger than expected) have led authors to propose that electron states (due to disorder, defects or impurities) are present in the forbidden gap of the insulator but are localised at the surface of the material [27]. Different models have also been proposed. Some of them assume that the final state is out of equilibrium. At the opposite, others propose that the materials are in equilibrium after contact. From their experimental data on oxide, Saint Jean and co-workers [29] have concluded that the charge transfer is related to an equilibrium final state. They have also shown that the occupied states in the gap are "surface states" even if these states are spread in the insulator (here $Al_2O_3$) over a distance larger than 50 nm. These "surface states" are supposed to be separated from the purely bulk states, since the charge density varies linearly with the contact potential (an exponential variation would be expected in the case of complete equilibrium with the bulk states).

In the case of polymers, the charge acquired by a particular material appears to be related to its chemical nature. Hence, the electrification of organic polymers seems to be strongly influenced by the grafted (donor or acceptor) groups onto the polymer backbone [35, 36]. Finally, it is worth noting that charge transfer is often limited by a feedback mechanism (such as back-tunnelling) during the separation of the metal and the insulator surface (oxides, polymers...), or by loss of charge by ionisation of the atmosphere when experimenting under normal atmospheric conditions.

*(iii)For insulator-insulator contact*, charge densities are in the same range as for metal-insulator contact. For instance, charge densities from 5 to 20 $mC/m^2$ have been measured for simple static contact between silica and mica [9]. All three of the above mentioned mechanisms (electrons transfer, ions transfer and material transfer) may occur [37]. However, the main mechanism for charge transfer is also probably electron transfer. Indeed, different authors have demonstrated [27, 34, 35] that the charge transfer between two dielectrics (polymers, oxides...) can be correlated to the charge acquired by each of them when contacting a metal. This has led to the controversial concept of triboelectric series, which aims to rank insulators regarding to their charging efficiency. A material ranked higher in the series, near the positive end, should charge positive when touched or rubbed with a material lower down, towards the negative end [32].

Another question concerns the possible difference between the generation of charges during static contact and friction. Generally, the total charge transferred during friction appears greater than during static contact. This has been explained by the increase of actual contact area during friction [27], so that the charge density is almost constant. Increase of temperature at high speed of sliding can also be implied. The major difference between friction and static contact is revealed by the possibility, for nominally identical insulators, to charge during rubbing and not during simple contact. This can be explained by the asymmetry introduced by the sliding movement and does not contradict the idea of electron transfer.

The penetration depths of the charges generated during these electrification has also to be considered. The experimental results present large disparities in the reported values of penetration depths [29, 38]. However, two tendencies have been evidenced. First, the penetration depth seems to be very sensitive to the resistivity of the insulator. For instance, in polymers, large penetration depths around 1μm have been reported on poly(N-vinylcarbazole) [39] which has a high charge mobility, and very small penetration depths have been measured on fluoropolymer [38]. In the last case, the value of the penetration depth is no more than a few nanometers, which is in the range of electron tunnelling, and no conduction seems to take place. For oxides, the penetration depth in $Al_2O_3$ is around 100 nm [29] and 30 nm for $SiO_2$ films [30]. Second, there are some experimental evidences that repeated contacts (and/or rubbing) allow charges to penetrate in the bulk of the sample for example by electrical conduction [38].

## 3.3 Conduction : the theory of polaron

The question of penetration depth implies the knowledge of the conduction in insulating materials. This conduction is related to the presence of defects in the material, on which charges can be trapped. The theoretical problem of conduction in insulators is described by the physic of the space charge and of the polaron in the case of inorganic crystallised dielectric materials [40-44].

As proposed by Landau, an electron moving in an insulator interacts with the polarisation field of the medium. Then, the polarisation around the charge produces its progressive enclosing in a potential well. Indeed, assuming that polarisation charges are concentrated in a ring at a distance $r_p$ from the electron, the potential energy of the system is expressed as :

$$W_p = -2E_R[1/\varepsilon(\infty) - 1/\varepsilon(0)][r_B/r_P] \quad (3)$$

where $E_R = 13.56$ eV is the Rydberg energy, $r_B = 0.05$ nm is the Bohr radius and $\varepsilon$ is the dielectric constant (respectively static $\varepsilon(0)$ or high frequency $\varepsilon(\infty)$ dielectric constant). This potential energy is negative because $\varepsilon(\infty) < \varepsilon(0)$ and the electron is coupled to the medium, the radius $r_p$ depending on the strength of this coupling. In the case of $TiO_2$, $W_p = 0.6$ eV $= 9.6 \cdot 10^{-20}$ J.

Polaron is considered as a quasi-particle including the charge and the polarisation around it (cf. fig. 4 (a)).

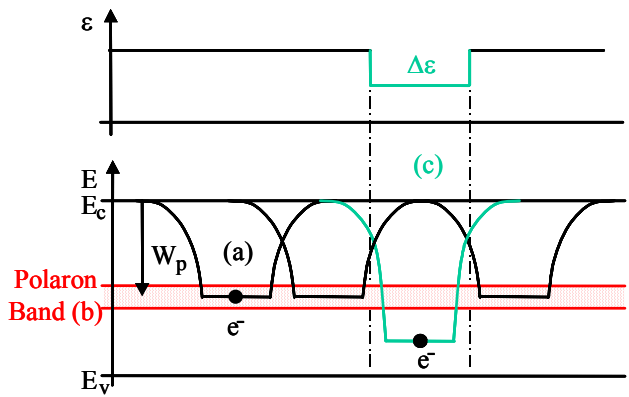

*Figure 4 :*
*Illustration of the band diagram related to the conduction and the trapping of the quasi-particle formed by the charge surrounded by the polarisation charge (also called polaron) in a dielectric :*
*(a) the <u>potential wells</u> (in black) associated with polaron sites,*
*(b) the <u>polaron band</u> (in red) produced by the overlapping of polaron orbitals. It allows the displacement of the polaron in the lattice. Its position in the band gap $W_p$ is given by equation (3).*
*(c) the <u>trapping effect</u> (in green) which is produced on a polaron site where there is a local decrease of the local electronic polarisability ($\Delta\varepsilon$).*

Polaron has hence an heavy effective mass, strongly dependent on the radius $r_p$. In the insulators, only small polarons are to be considered, their effective mass is very large and the radius $r_p$ is so small, that it can be smaller than the interatomic distance ($r_p = 0.11$ nm for $TiO_2$). In this case, a narrow polaronic conduction band appears due to the overlapping of the wave functions on adjacent sites (cf. fig. 4 (b)). Depending on the temperature, two conduction mechanisms are present in these insulators. At low temperature the polaron moves in the narrow band whenever at high temperature, conduction by hopping between sites is thermally activated. At room temperature, the mobility is around 10 $cm^2/V.s$, which is one or two order of magnitude smaller than in the case of semiconductors. Suppose now that there is a slight variation in the dielectric function due to the presence of disorder or defects (cf. fig 4 (c)) [45]. The binding effect of the charge is reinforced, and the trapping of the charge occurs. The defects responsible of this trapping of polaron are hence defects associated with local variation of polarisability. This is directly related to the well accepted idea that trapping in insulator is due to defects. In inorganic materials such as ceramic, the variation of polarisability is coming from point and extended defects such as vacancies [46], segregation, grain boundaries [47]…For crystalline polymers, the structural defects are so numerous that they give a quasi continuous distribution of polaron traps. This approach of the conduction proposed by Blaise [45, 48] differs from the more classical one [49], because it does not involve exchange of electron or hole between the rigid band structure and discrete levels localised in the band gap. In this new approach, the trapping-detrapping process is entirely restricted to the polaron band resulting from the polarisation properties of the medium.

At the end of this section, it is worth noting that some bad results on the triboelectric series can be explained by the previous conduction and trapping theory. Actually, the idea of ranking insulators in triboelectric series has more or less failed, because a given substance does not occupy an invariable position in the series. For instance, heating a sample, or polishing it, may cause it to move up or down in the series [32]. Therefore, because the defects (considered as traps for the charge carriers) are responsible of the quantities and the nature of the exchanged charges between materials (insulator-metal or insulator-insulator contact), this apparent failure of triboelectric series seems quite logical. Thermal treatment [50] or polishing [51] modifies the number and (or) the nature of defects present in insulators so that its charge transfer capability will be modified and hence its position in the triboelectric series.

**4. Role of the electrical charges on friction and adhesion**

In section 3, we have recalled how electrical charges can be generated during friction or static contact. We have also shown that these charges can move and be trapped inside the material. The brief overview in section 2 on the contribution of electrical charges to the surface forces has evidenced that until now, the possible effects of the trapped charges on the interfacial energy during friction of insulators has been neglected or could not be properly described using the existing models. In the present section, based on our own experimental results, we are going to demonstrate that interaction energy during friction may depend markedly on the effect of the trapped charges.

**4.1 Scanning electron microscope mirror method (SEMM)**

In order to characterise the trapped charges, an original method called the SEMM method, has been developed [52]. This method allows us to know the ability for an insulator to trap charges. It has been validated in studies on space charge associated to breakdown of insulators [48]. The principle is the following. The insulating sample is irradiated in a SEM at high voltage (30 keV). The total amount of charges injected in the sample is $Q_i$. Negative charges trapped in the insulator during the injection, $Q_t$, produce an electrical field in the vacuum chamber of the SEM. If the sample is observed later, with a lower energy electron beam (100 to 1000 eV), the electrical field can be strong enough to deflect the electrons in the same manner as a convex mirror does with light (trajectories 1 and 2 on Fig. 5-a).

As a consequence, a mirror image (Fig. 5-b) is given on the screen which displays a distorted view of the SEM chamber. The amount of trapped charge $Q_t$ is calculated using an electrostatic law. On one side, the trapping yield, which is ratio between the trapped charge $Q_t$ divided by the injected charge $Q_i$, characterises the capacity of the material to trap electrical charges. On the other side, the variation of the trapping yield as a function of the temperature gives access to the depth and the activity of the traps which depends on temperature.

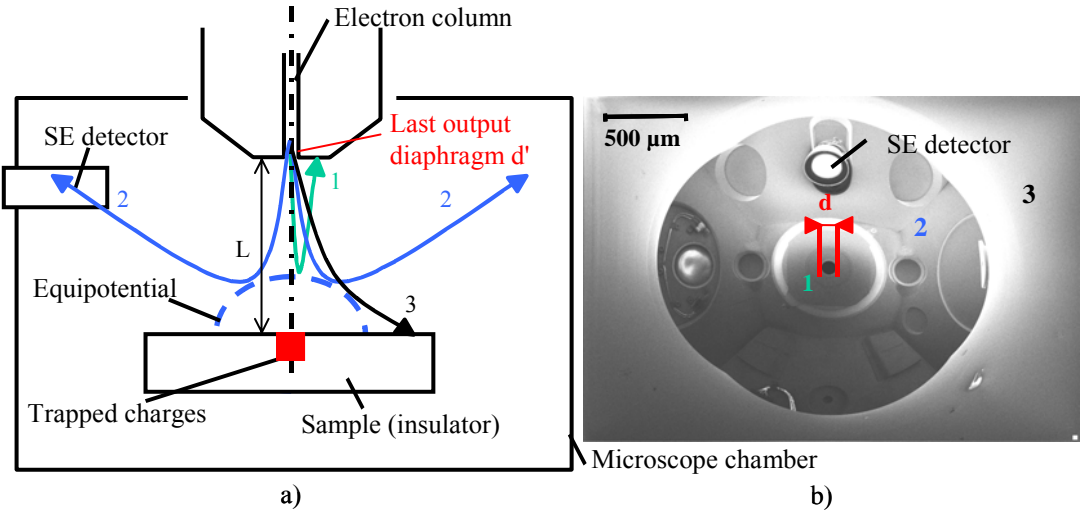

*Figure 5 :*
*SEM Mirror Method : First, charges are injected and trapped in the sample($Q_t$ in red) through a high energetic electron beam. Second the sample is observed with a low energetic electron beam, whose electrons are deviated by the electrostatic field due to the trapped charges (equipotential in blue).*
(a) -Electron 1 go back in the electron beam and give no image which gives the central black disk in (b). The electrostatic law relating the real diameter d' of the last output diaphragm and the apparent one d measured on (b) is the following : $\frac{1}{d} = \frac{4L}{d'} \frac{V_s 2\pi\varepsilon_0 (\varepsilon_1 + 1)}{Q_t}$ with L the working distance of the SEM, $V_s$ the surface potential and $\varepsilon_r$ the dielectric constant of the sample. By plotting d versus $V_s$, the trapped charges $Q_t$ can be deduced.
- Electron 2 are reflected by the microscope chamber and give an image of its walls.
- Electron 3 are just deviated by the trapped charge and give a distorted image of the sample.
(b) Distorted view of the SEM chamber after trapping of charges.

## 4.2 Experimental results

Main of our results have been obtained on sapphire (single crystal of alpha-$Al_2O_3$) during friction tests performed on a flat on flat reciprocating tribometer. We have evidenced a strong correlation between the ability for an insulator to trap charges and its tribological properties. These results can be summarised as following. In a first study [53], it has been shown that for pure sapphire heat-treated at 1500°C, no measurable charging effect can be detected at room temperature before or after static contact. On the opposite, after five friction cycles, charging capacity of the sample was measurable by SEMM method inside and outside the friction track. This has revealed that, as soon as sliding occurs, defects are created everywhere in the sample and this has been correlated to the increase of the friction coefficient as a function of the cycle number. Naturally, the pre-existing sites of trapping, present before friction, have a role on the evolution of the friction coefficient. This fact has been evidenced by modifying the defect population of sapphire by X-ray irradiation [54]. For X-Ray irradiated samples, the friction coefficient has reached a steady state value around 0.8. When, for non-irradiated sample, the steady-state value was only about 0.3. The SEMM study of these two samples has revealed that the irradiated samples have acquired a very large ability to trap charges.

*Model of adhesive and load controlled contact*
To analyse the friction tests, the model initially described by Tabor [55, 56] was chosen, because the load used during the friction test led to values of the external pressure in the range of the values of the adhesive pressure (0.2 MPa to 0.3 MPa). At the microscopic level, the stabilised friction force $F_x$ arises from three contributions: the ploughing of surfaces by the asperities, the adhesion and the film shearing. For ceramics with very low roughness, the ploughing contribution can be ignored and the friction coefficient µ is given by $\mu=F_x/F_z=\tau A/F_z=\tau/p_{ext}$, where $\tau$ is the mean shear strength, A is the real area of contact and $p_{ext}$ is the mean pressure of the contact.

Then, the model considers the energy expended in moving the two surfaces. Assume two surfaces which asperities are periodic and interpenetrated (cf. fig. 6), each asperity occupying an elementary surface a. If the two surfaces are separated to infinity, they would have free surface energies γ and the interaction energy W. Between two asperities, the interaction energy $W_a$ would be equal to 2aγ. In the energy expended in dragging one asperity over it neighbour (lateral movement $\Delta x$), it is assumed that only a small fraction $\zeta W_a$ is involved, because the distance required to initiate motion is only $\Delta z$. On an empirical basis, the fraction ζ is usually taken equal to 1/30.

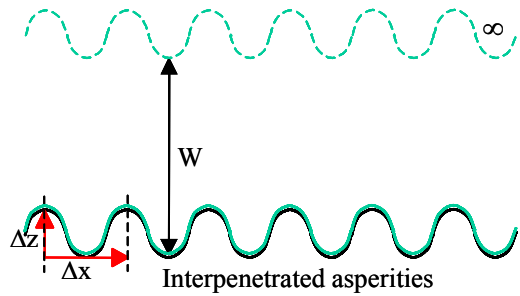

*Figure 6 :*
*Microscopic representation of the interpenetrated surfaces in contact ($\Delta z$=height of asperities, $\Delta x$=width of asperities)[56]. W interaction energy for infinite separation.*

Let us consider now the two contributions of the mean shear strength separately. Assume firstly that the external pressure is negligible compared to the adhesive contribution. The shear strength necessary for sliding of an elementary surface is $\tau_{ad}=F_x/a$. Thus, the condition necessary for sliding is obtained by equating the two energies : $\Delta x \cdot F_x=\zeta W_a=\zeta 2a\gamma$, and the expression of adhesion contribution to the shear strength is : $\tau_{ad}=2\zeta\gamma/\Delta x$. Assume secondly the opposite case, with the magnitude of the externally applied pressure higher than the adhesive pressure. The condition for sliding is also obtained by equating the two energies : $\Delta x \cdot F_x=\Delta z \cdot F_z$. Thus, we obtain the expression of the external pressure contribution to the shear strength : $\tau_{ext}=F_x/a=\alpha p_{ext}$, where $p_{ext}=F_z/a$ is the applied pressure on an elementary surface and $\alpha=\Delta z/\Delta x$ is related to the sub-microscopic surface roughness.

Considering the total shear strength in the contact, the friction coefficient is:

$$\mu=\frac{\tau_{ad}}{p_{ext}}+\alpha \qquad (4)$$

Now, whatever the normal load is and if there is no wear during the friction test, roughness is not modified. We thus assume that the α coefficient is independent of $p_{ext}$, and equation (4) shows that the friction coefficient has a linear dependence with the inverse of the external pressure. The slope depends on adhesion while the constant α is characteristic of the surface geometry.

*Friction test results*

Results obtained by varying the applied pressure during friction on a reference alumina (annealed during 4h at 1500°C in air), confirm the linear dependence with $1/p_{ext}$ (cf. fig. 7). Then, different treatments (thermal annealing and UV irradiation) have been made, which do not modify the surface roughness (α remains almost constant, α= 0.1). For thermal treatments, the linear dependence of the friction coefficient versus the inverse of the external applied pressure is still obtained but the slope decreases when the annealing temperature increases (cf. fig7 tab.1). Similar results have been observed for UV irradiated samples (cf. fig. 7 and tab. 1). This variation of the slope accounts for the variation of the adhesion term due to these treatments.

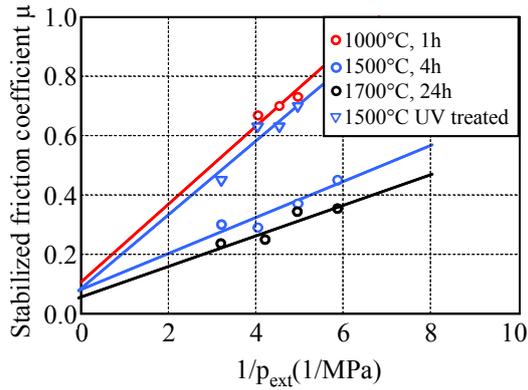

*Figure 7 :*
*Stabilized friction coefficient μ of pure sapphire as a function of the inverse of the external applied pressure (flat on flat friction experiments, contact area S=78.5 $mm^2$). Different thermal treatments have been applied (circles) and for 1500°C, an additional UV irradiation has been performed before the friction test (triangles). Note that the friction coefficient is consistent with Tabor's model : $\mu = \tau_{ad}/p_{ext} + \alpha$ . where $\tau_{ad}$ is the adhesive part of the shear strength and $\alpha$ is a constant depending on the roughness of the surface (almost the same in these cases). The slope depends on the different treatments so that it can be deduced that the treatments modify the adhesion contribution to the friction coefficient. This has been related to the variation of the trapping ability of the material with the different treatments.*

*SEMM results*
However, the dielectric study of these different treated samples has revealed an important difference in their respective ability to trap charges (cf. tab.1). Indeed, it appears that the ability for sapphire to trap charges decreases with the annealing temperature and in contrast, increases with the UV irradiation. The cathodoluminescence study [46] of these samples has allowed us to propose an explanation of the trapping results as a function of the electronic configuration and the number of point defects in sapphire such as oxygen and aluminium vacancies.

| Treatment | $\tau_{ad}$ (MPa) | $\Pi = Q_t/Q_i$ |
|---|---|---|
| 1000 °C | 0.131 | > 0.1 |
| 1500 °C | 0.060 | 0.02 |
| 1700 °C | 0.051 | 0 |
| 1500 °C + UV irradiated | 0.121 | 0.1 |

*Table 1. Adhesion contribution $\tau_{ad}$ to the shear strength and trapping yield ($\Pi = Q_t/Q$) determined at room temperature as a function of the sapphire treatments*

*Interpretation*
For thermal treatments, when the annealing temperature increases, the number of active oxygen vacancies decreases, so that the quantity of trapped electrons (generated during friction) also decreases. According to the image charge theory, this could partially explain why the adhesion contribution to the friction coefficient decreases, when the annealing temperature increases. In a similar way, the UV irradiation leads to a transfer of electrons from O-vacancies to Al-vacancies [46]. As a consequence, more O-vacancies become able to trap electrons and the number of trapped electrons (generated during friction) becomes higher than for non-irradiated sample, leading to a higher contribution of the adhesion to the friction coefficient.

Finally, as proved by image charge interactions (cf. § 2.3), adhesion energy of the interfacial films present in the contact depends on the charges located near by the surface. Consequently, a change in defect configuration induces charge modification and thus adhesion modification.

### 4.3 Discussion

In the previous sections, qualitative evidences have been given concerning the effects of charges on surface forces, on adhesion and on friction. It has also been shown that surface force measurements [9] can be a relevant technique to study these effects.

But some fundamental questions remain, which are given below :

- What is the quantitative part of the electrostatic forces in surface forces compared to the other contributions ? How does it depend on solid separation and/or contact pressure ?
- What is the role of charges in adhesion, in friction, in wetting phenomena and how can this be quantified ? The contribution of charges on free surface energy is still to be quantified.
- What are the mechanisms of charging phenomena, of charge trapping/detrapping and of charge movement ? Is this related to stresses in the material under pressure or during rubbing ?

In order to go further on these points, one way can be to combine accurate surface force measurements with a precise knowledge of the dielectric properties and of the electrostatic state of the material. That is why we propose, in a future work, the coupled use of the SEMM method and of SFA [57]. Sapphire will be used because it can be smoothed enough to allow us to perform surface force measurements, and its trapping ability has been already extensively studied previously. The initial stress state will be controlled by adequate thermal treatments. UV irradiation and mechanical solicitations, performed with SFA used as nano-tribometer for example, combined with in-situ surface force measurements will be used in the aim to correlate the mechanical and physico-chemical properties to the dielectric characteristics of the material.

### 5. Conclusion

The main points to emphasize are these. First, charges occur easily during simple contact or friction, in polar or non-polar environment. It does not occur solely at the surface but also in the bulk by charge trapping, which induces long distance effects, according to space charge physics. Second, charges can play a major role in adhesion energy as proved by both experimental results and image charge effects. Third, interaction energy during friction may depend markedly on the effect of the trapped charges and, consequently on the presence of structural defects pre-existing or created during friction.

It would inappropriate to end without mentioning the role of these charges in the mechanical or electrical degradation of insulating materials. In fact, charge localisation induces storage of polarisation energy by which chemical change or defect production is achieved. Particularly mechanical stress may not be the only driving force for dislocation growth or motion since the dislocation in ionic solids can be charged and the internal electric field can have a direct effect on their movement [44]. In this way, dislocation growth observed during indentation [51] or friction [58] can be explained, as for breakdown [59], by trapping of charges on point defects leading to the formation and motion of dislocations both inside and outside the stressed zone. The localisation and then release of polarisation energy, especially along mesostructure (grain boundary and dislocation network...), can explain, for example, severe wear of ceramics [60].